# Flexible and disposable paper and plastic-based gel micropads for nematode handling, imaging and chemical testing

Zach Njus,[1] Taejoon Kong,[1] Upender Kalwa,[1] Christopher Legner,[1] Matthew Weinstein,[1] Shawn Flanigan,[2] Jenifer Saldanha[3] and Santosh Pandey[1*]

[1]Department of Electrical and Computer Engineering, Iowa State University, Ames, IA 50011, USA.

[2]Department of Mechanical Engineering, Iowa State University, Ames, IA 50011, USA.

[3]Department of Genetics, Development and Cell Biology, Iowa State University, Ames, IA 50011, USA.

* E-mail: pandey@iastate.edu

**Abstract:** Today, the area of point-of-care diagnostics is synonymous with paper microfluidics where cheap, disposable, and on-the-spot detection toolkits are being developed for a variety of chemical tests. In this work, we present a novel application of microfluidic paper-based analytical devices (µPADs) to study the behavior of a small model nematode, *Caenorhabditis elegans.* We describe schemes of µPAD fabrication on paper and plastic substrates where membranes are created in agarose and Pluronic gel. Methods are demonstrated for loading, visualizing, and transferring single and multiple nematodes. Using an anthelmintic drug, levamisole, we show that chemical testing on *C. elegans* is easily performed because of the open device structure. A custom program is written to automatically recognize individual worms on the µPADs and extract locomotion parameters in real-time. The combination of µPADs and nematode tracking program provides a relatively low-cost, simple-to-fabricate imaging and screening assay (compared to standard agarose plates or polymeric microfluidic devices) for non-microfluidic, nematode laboratories.

**INTRODUCTION**

Devices fabricated on low-cost, flexible and disposable substrates such as paper, plastic, and textiles have gained considerable interest with applications in resource- and power-limited settings.[1-3] The acronym µPADs (i.e. microfluidic paper-based analytical devices) represents a class of devices where microfluidic networks are made in paper substrates by patterning hydrophobic and hydrophilic regions within them.[4,5] Here the hydrophobic barriers restrict the fluid flow within pre-specified areas of the cellulose paper while the fluid spreads in the hydrophilic regions by capillary action. Today, this simple concept of designing fluid handling using hydrophobic/hydrophilic patterns has evolved into a promising µPAD technology that has the potential to revolutionize on-field screening of disease biomarkers.[1,6]

It was realized early on that the major drivers of µPAD technology are the material/processing cost per device and the capability of mass manufacturing.[6] As such, a variety of low-cost, printing techniques were explored on paper substrates that require minimalistic handling steps. Wax printing is probably the most direct way of fabricating µPAD structures;[7-9] however other µPAD fabrication methods have been demonstrated that require additional chemical processing or machining steps (e.g. inkjet printing, wet etching, laser treatment, screen printing, and flexography).[10-16] Wax printing of µPADs is also appealing because it involves simple fabrication steps (i.e. printing using a commercial wax printer and baking on a hot plate), uses inexpensive and chemically-inert materials (i.e. paper and wax), and is environmentally friendly (i.e. no use of organic solvents).[4-6] There is no need for clean room facilities and devices can be fabricated on-site by personnel with minimal training. Ongoing research directions in µPAD fabrication methods are exploring the use of bioactive substrates with embedded and functionalized nanoparticles, printing biomolecules directly on testing zones, integrating multiple substrate layers into functional 3D structures, and incorporating control components to regulate fluid flow in both lateral and longitudinal directions.[17-21]

A number of µPAD applications have been reported in areas of medical diagnostics, environmental monitoring, and food safety.[1,4,11,16,20] The general experimental process flow is as follows: an unknown sample reaches the detection zone containing known reagents that triggers a certain biochemical reaction. This, in turn, generates an output signal that is detected by optical, fluorescent, chemiluminescent or electrochemical techniques.[6] Optical or colorimetric detection methods are the popular choice where low-cost, portable, hand-held detectors can record visual changes in the biochemical reaction. Various clinically-relevant analytes (e.g. glucose, protein, and alanine aminotransferase) have been analyzed from biological fluids (e.g. blood, serum or urine) using enzymatic, color-based detection with the possibility of imaging and transmitting the results by telemetry (using smartphones, cameras or scanners).[22-24] Colorimetric detection of protein and DNA-based biomarkers have been demonstrated where antibodies immobilized on chemically-modified µPADs captured the targeted antigens, producing a distinct color change in the colorimetric reagent.[24-26] Attempts have been made to improve the sensitivity and lower the colorimetric detection limit by using chemically-modified paper to reduce surface fouling,[25] incorporating functionalized nanoparticles with high molar absorptivity to select biomarkers,[26] and modifying the smartphone's camera system to permit testing in ambient lighting conditions.[27]

As new and elegant applications of µPADs emerge within the realm of molecular diagnostics, it is intriguing to consider the possibility of using paper-based devices for small animal studies, particularly those involving nematodes or worms. *Caenorhabditis elegans* is an important, small model nematode pivotal to experiments on gene regulation, metabolism, ageing, cell signaling, chemical screening, drug discovery, and space flight.[28-32] Compared to their mammalian counterparts, *C. elegans* are well-suited for high-throughput, large scale biological experiments because of the ease to culture on a diet of *Escherichia coli*, ability to grow from an egg to an adult within three days, and capacity to produce over 300 progeny.[28,29] In this context, polymer-based microfluidics has emerged as an enabling technology for *C. elegans* research where on-chip automation has streamlined the steps of worm capture, immobilization, transport, screening, sorting, and/or tracking.[33-36] For instance, in the field of pharmacology and drug discovery, it is possible to trap individual worms within discrete liquid droplets in a microfluidic chip and simultaneously screen them for a range of drug concentrations.[35,36] In the field of neurobiology, individual worms can be rapidly immobilized in microfluidic chambers where laser nanosurgery can be performed on their axons.[37] These and many more examples[38-41] show how microfluidics can be leveraged to physically or chemically manipulate worms' microcosm, along with large-scale genetic and behavioral screening to predict the role of underlying genes affecting an observed behavior.

However, technological advancements in polymeric microfluidics are generally limited to engineering disciplines and have yet to reach a majority of nematode biology laboratories.[41] One hurdle in the wide-scale adoption of polymeric microfluidics by worm biologists is the access to microfabrication facilities and engineers to iteratively design, fabricate, and test the microfluidic devices. Most of the worm chips described above necessitate the use of multi-layer device fabrication, sophisticated pumps and valve operations, and advanced programming skills to control and synchronize system operations.[37-41] The cost of chemicals and consumables in microfluidics can be expensive to sustain for repeated experiments when a regular inventory of devices is needed to test the biological hypotheses. Because of the above reasons, routine *C. elegans* experiments are often performed on petri dish and well plates that offer design simplicity, easy accessibility, experimental flexibility, standardized manufacturing, and disposability.

Here we attempt to gauge the applicability of paper-based microfluidic technology to the study of *C. elegans*. The following reasons justify the need of paper-based microfluidics for worm biology. Firstly, paper is a ubiquitous, easily-available, and disposable substrate that is far cheaper than standard polymers used in conventional microfluidics.[5] Secondly, the fabrication procedure of paper-based devices is straightforward and a variety of cost-effective printing methods can be easily adopted in biological laboratories.[7] Thirdly, there is no requirement of external instrumentation as paper-based devices are designed to work passively.[4-9] Fourthly, successful attempts have been made to co-culture different cell populations on gel spots in paper microfluidics where the benefits of paper-based *in vitro* cell culture assays are outlined.[42-43] As a tradeoff, paper-based devices do not have the resolution of device dimensions obtained in polymeric microfluidic devices.[44] Thus it is possible to utilize the benefits of paper-based devices for specific applications in worm biology where low cost and operational ease are important while the device dimensions can be compromised upon.

In this work, we present an open platform comprising µPADs on paper and plastic substrates to conduct worm experiments. Initially, the methods of device fabrication are described where the paper and plastic substrates are machine cut, and agar or Pluronic gel membranes are suspended on the excised regions of the substrates. The suspended membranes serve as robust platforms for imaging worm behavior over extended time periods; both at single and population levels. Next, we show schemes for chemical testing where the worm is subjected to a drug and its dose response is characterized using a custom software program. Transfer of single and multiple worms from one µPAD membrane to another membrane or a petri dish is relatively simple and direct here without the need for picking individual worms. A method of z-stacking is demonstrated where two worm populations, each on separate vertically-stacked membranes, are imaged simultaneously. Taken together, we propose agar or Pluronic gel µPAD membranes suspended in paper and plastic substrates as a feasible platform for worm handling, imaging, and chemical testing.

**MATERIALS AND METHODS**

**Device Fabrication**

Two substrates are used for fabricating the open microfluidic devices: paper and plastic. The fabrication steps are illustrated in Fig. 1. In general, both paper and plastic substrates are made hydrophobic by the use of a wax barrier or tape. Circular areas are excised in the substrates that eventually support a thin membrane of agarose or Pluronic gel. The membrane forms the open µPADs on which *C. elegans* locomotion behavior will be observed and imaged. The boundary of the circular areas ensure that worm movement is always restricted to the µPAD membrane and within the field of view, which is not possible on agarose plates where worms can wander outside the field of view.

The paper-based µPAD is fabricated using Whatman™ chromatography paper (Sigma Aldrich™) as shown in Fig. 1a. Initially, the device design is drawn in Adobe Illustrator™ and printed on the chromatography paper using a wax printer (Xerox Colorcube™). The paper is placed on a hot plate at 130°C with the wax pattern facing upwards. After 15 seconds, the paper is removed from the hot plate and allowed to cool at 23°C. A biopsy punch (Miltex™) is used to excise a circular area within the wax pattern, which is then lowered onto a 100 µL droplet of 2% liquid agar (maintained at 80°C). When the interior of the wax pattern has come in contact with the liquid agar, the paper is then lifted off the droplet. This creates a suspended membrane of liquid agar in the circular excised area (Fig. 1a(v)). This agar membrane is hydrated with 15 µL of M9 buffer and placed in a humidified petri dish sealed with parafilm.

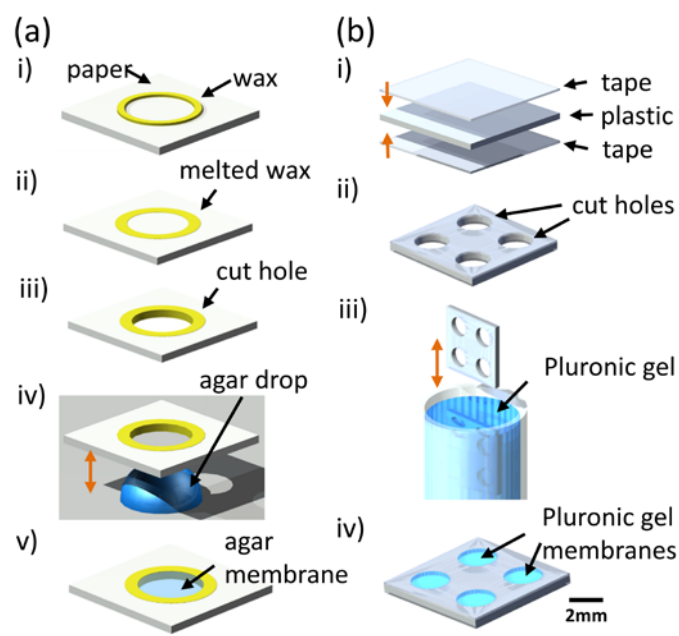

**FIG. 1.** Fabrication of suspended gel membranes on paper and plastic substrates. (a) Paper-based µPADs: A wax pattern is printed on chromatography paper and melted on a hot plate to create a hydrophobic barrier (i-ii). A circular pattern is excised using a biopsy punch to create a cut hole (iii). The paper is lowered onto a 100 µL droplet of 2% liquid agar and gently lifted, resulting in an agar membrane in the cut hole (iv-v). (b) Plastic-based µPADs: A plastic sheet is taped on both sides and multiple circular patterns are excised using an automated design cutter to create cut holes (i-ii). The excised plastic sheet is vertically immersed into a 11.5% solution of Pluronic gel and removed, resulting in suspended membranes of Pluronic gel within the cut holes (iii-iv).

The plastic-based µPAD is fabricated using 8.5″ × 11″ transparency sheets (Staples Inc.™) and packaging tape (Scotch™) as shown in Fig. 1b. Initially, the device design is drawn in Silhouette Cameo Studio™. Tape is applied to both sides of the transparency sheet and multiple circular patterns are excised using a Silhouette Cameo™ electronic cutter. The transparency sheet is vertically dipped into a liquid solution of 11.5% Pluronic gel and then lifted upwards out of the solution. This creates Pluronic gel membranes in the excised areas of the transparency sheet (Fig. 1b(iv)). The gel membranes are hydrated with 15 µL of M9 buffer and placed in a sealed, humidified petri dish.

**Experimental Setup**

Wild-type *C. elegans* are grown on standard Nutrient Growth Media (NGM) agar plates with *Escherichia coli* OP50 at 20°C following previously published methods.[28] The petri dish housing the open µPADs (paper-based or plastic-based) is placed on the stage of a Leica MZ16 stereozoom microscope (7.1× to 230× magnification range) at an ambient temperature of 22°C. Two methods are used to put worms on the membrane devices: pipetting worm droplets and picking/insertion of worms. In the pipetting approach, individual L4-stage worms are collected in a droplet of M9 buffer, and pipetted directly onto the agar membrane (Fig. 2a). The worms initially appear to swim in the liquid droplet but, as the excess buffer is absorbed by the agar, the worms start to crawl on the agar surface. In the picking/insertion approach, individual worms are picked by a platinum wire pick. The wire pick is plunged through the Pluronic gel and extracted out, leaving behind the worms within the gel (Fig. 2b). Based on our trials, the pipetting approach is suggested for paper-based devices consisting of agar membranes, whereas the picking/insertion approach is well-suited for plastic-based devices consisting of Pluronic gel membranes. This is because any physical contact from the wire pick may produce undesired irregularities on the agar membrane surface, while the Pluronic gel membrane is insensitive to any physical contact or penetration by the pick.

After placing the *C. elegans* on the µPADs, the worms are visually observed under the microscope to ensure that they appear healthy and mobile. After a wait time of around 3 minutes, worm movement is recorded for the next 3 minutes at 10 frames per second using a QImaging 12 bit color camera. In the case of drug testing, 5 µL of levamisole solution (in M9 buffer) is added on top of the membrane and the worm responses are recorded. No ethics approval is required.

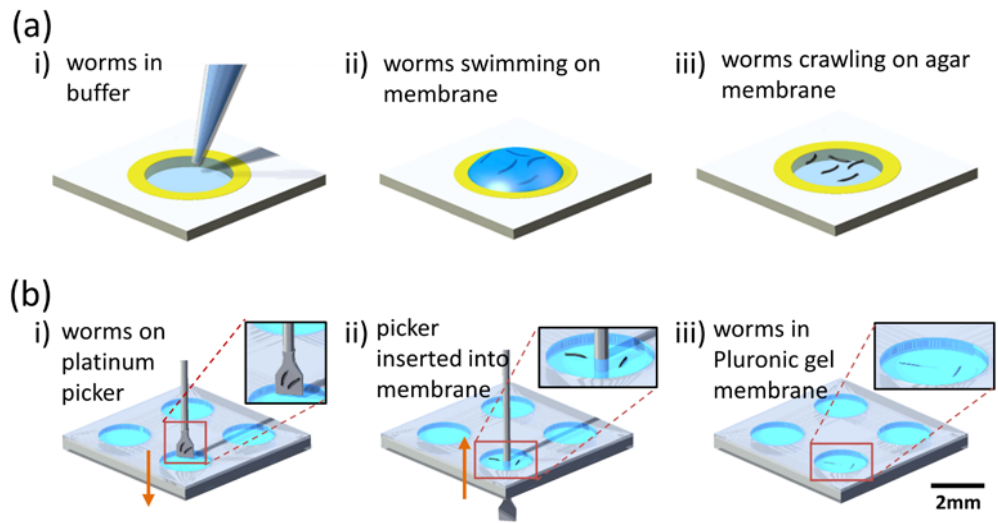

**FIG. 2.** Two methods of placing *C. elegans* on suspended gel membranes. (a) Pipetting worm droplets on paper-based agar membranes: A liquid droplet containing worms is pipetted onto the agar membrane and the worms swim in the droplet (i-ii). The excess liquid is allowed to be absorbed into the agar and the worms appear crawling on the agar surface (iii). (b) Picking and inserting single worms in plastic-based Pluronic gel: A platinum wire pick is used to select and pick single worms, which is then plunged through the Pluronic gel membrane (i-ii). Upon extracting back the pick, the worms remain within the Pluronic gel (iii).

**Data Collection and Analysis**

The videos are run through a custom tracking program written in Matlab (Mathworks Inc., Natick, MI) that identifies individual worms and tracks their positional coordinates through a series of image processing steps on the raw data (Fig. 3a). Initially, the worm-tracking program identifies circular membranes present in the fabricated µPADs (Fig. 3b). This step helps to limit the area observed, thereby minimizing the false identification of worm-shaped objects and decreasing the processing time. In order to identify the circular membranes, we use the Circular Hough Transform (CHT). The CHT is a popular algorithm for finding circular shapes in images due to its ability to handle occlusions (overlapping or connected areas) and non-uniform illumination. We screen and qualify pixels with high gradients as candidate pixels. These candidate pixels are then used to vote in a circular pattern of a user-specified radius range forming an accumulator array. Clusters that arise in the accumulator array correspond to centers of circles present in the image. Since the circles identified from the CHT algorithm are not perfect circles, the edges of the found circles need to be further refined. To refine the edges, the found circles are scaled to the original size of the image and used as an initial mask input to an Active Contour Algorithm (ACA). The ACA calculates gradient field lines, which point towards edges in the image. The edge pixels of the circles are then directed into these edges, thereby refining the edges of the found circle (Fig. 3b).

Once the circular membranes are identified, the worms need to be identified and segmented from the video frames (Fig. 3c). To accomplish this task, an averaging method is generally used where the background image is generated by averaging a number of video frames. Any moving objects automatically average out. The background image is subsequently subtracted from the video frames to identify the moving object. However, this averaging method is not capable to of detecting sedentary worms within a video as they become part of the background. To identify and segment both mobile and sedentary worms from the background, we use the local thresholding technique. A window size of 100 × 100 pixels is scanned over the frame image and any pixels that are less than 90% of the average brightness of the window are set to white. This local thresholding approach is able to detect all worms (both moving and sedentary) even in a non-uniform lighting background. Later, a morphological classifier is used to remove any segmented sections that are not shaped as worms (Fig. 3d). We defined sections as worm-shaped if the ratio of the perimeter length to the area of the section is in the range of 0.5 – 1 and if the area of the section was 200 – 300 pixels. These values are chosen for L4-stage *C. elegans*, and may be altered depending on the size and shape of a different nematode.

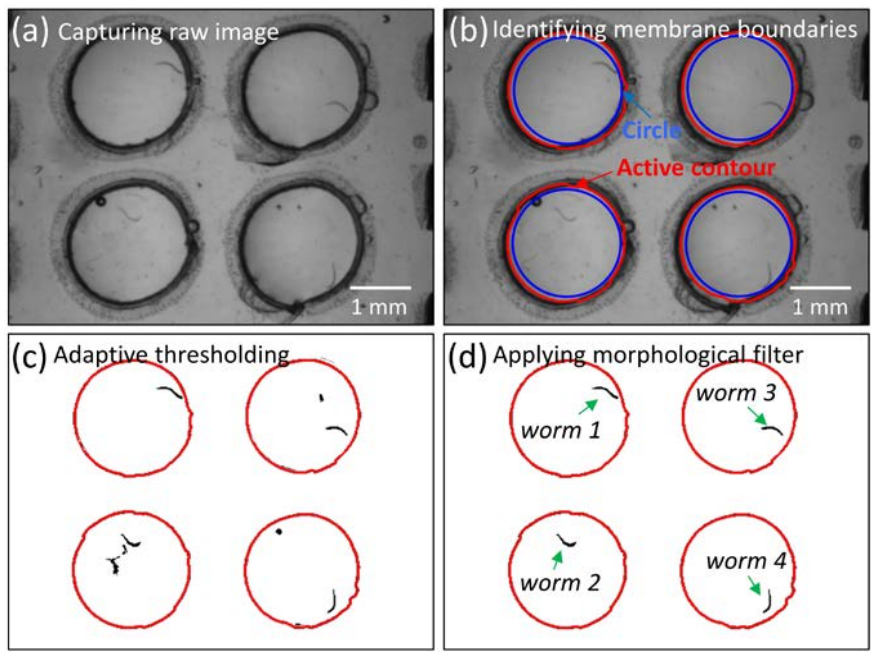

**FIG. 3.** Worm identification. a) Original image of the plastic-based µPAD with four suspended membranes, each housing a single worm. b) Circular membranes found by the Circular Hough Transform are illustrated in blue. Edges of the identified circles refined using the Active Contour Algorithm are shown in red. c) Sections are segmented out using an adaptive threshold technique with a window size of 100 x 100 pixels and a percentage of 90%. d) Using a morphological filter,

specific sections are removed that do not fit within a certain threshold (area of 200 – 300 pixels; perimeter to area ratio of 0.5 –1). Arrows point to the four worms identified after the last filtering step.

After the worms are segmented from the video, a minimal area-bounding box is fitted around each worm. The center point of the bounding box is considered the centroid location of the worm, which is tracked throughout the video and connected to form a complete path or track that the worm traced. If multiple worms occlude with each other, the area of the bounding box becomes greater than 300 pixels, and the paths are terminated until the worms resume their separate tracks. The track data is reported in the form of a Microsoft Excel workbook for individual worms. The velocity is calculated by measuring the distance moved by the worm's centroid between successive time points and dividing by the video frame rate. Eventually, each workbook consists of centroid locations and instantaneous velocity of the worm at every measured time point.

**RESULT AND DISCUSSION**

**Real-time worm tracking**

By the two methods described in Fig. 2, worms are put on paper- or plastic based μPADs. Fig. 4a(i) illustrates the motion of a single worm on the agar membrane of a paper-based device. Under a light microscope, the appearance of the agar membrane is clear while the paper substrate is opaque. The worm on the agar membrane is easily identifiable by the tracking program that produces the centroid locations of the animal throughout the duration of the 3-minute video (worm track colored in red). Fig. 4a(ii) shows the velocity of the centroid location for a representative worm that is crawling on an agar membrane with M9 buffer. Similarly, worms can be placed on plastic-based devices to track their movement. Fig. 4b(i) shows four plastic-based Pluronic gel membranes; each housing a single worm in M9 buffer. The appearance of worms swimming in the Pluronic gel membranes is not visually as clear as those crawling on agar membranes (shown in Fig. 4a(i)). Still the tracking program is able to identify the worms from the background and track the centroid locations of the four worms (colored in red, yellow, light blue, deep blue). Fig. 4b(ii) shows the centroid velocities of the four representative worms that are recorded swimming in the Pluronic gel membranes over a 3-minute time period. The position or velocity data can be used to calculate parameters related to behavioral traits such as attraction, aversion, toxicity, and social effects.[45-48]

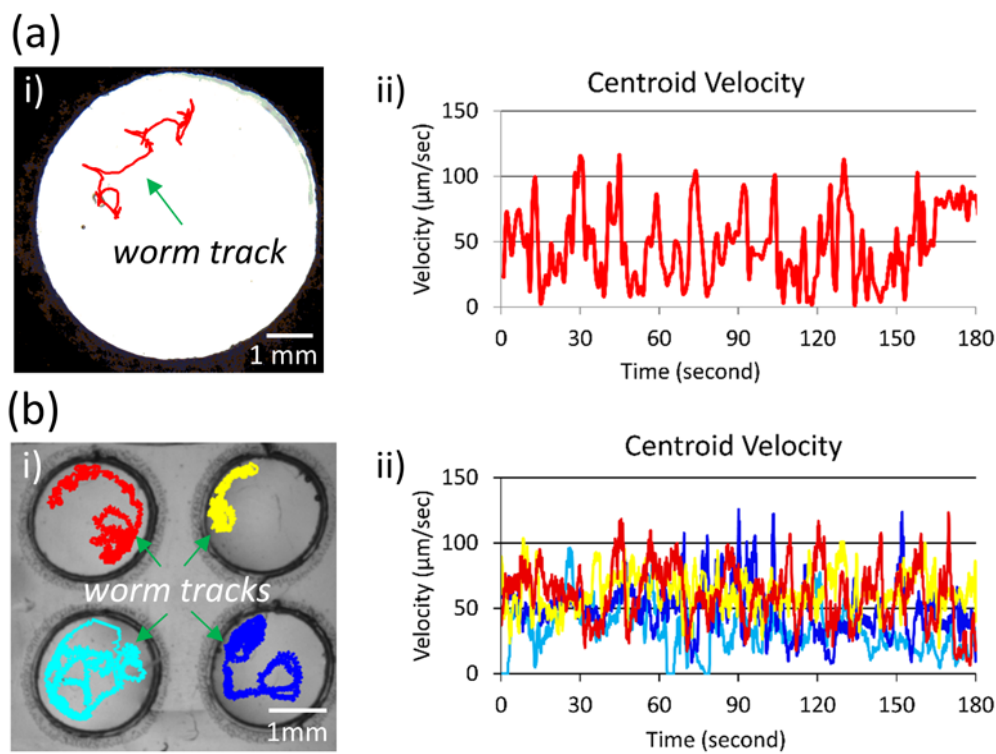

**FIG. 4.** Real-time worm tracking on paper and plastic-based devices. (a) The track of a worm crawling on a paper-based device is illustrated in red (i). Velocity plot of a representative worm's centroid location is shown over 3 minutes in M9 buffer (ii). (b) The tracks of four worms swimming in four separate plastic-based devices are illustrated (i). Velocity plots of the four representative worms' centroid locations are shown over the course of the 3 minute in M9 buffer (ii).

**Levamisole drug testing**

An important application of microfluidic devices in the *C. elegans* community is the screening of different chemicals and drugs for their relative toxicity.[45,49] Here we demonstrate the use of the membrane devices to test the *C. elegans* response to an anthelmintic drug, levamisole. A specific concentration of levamisole is prepared beforehand in M9 buffer and dropped on the suspended agar or Pluronic gel membranes. Individual worms are put on the μPAD membranes, allowed to acclimatize for 3 minutes, and then recorded for another three minutes. For every levamisole concentration, the tracking program produces the instantaneous centroid velocities of each worm which are then averaged for all the worms to give the average centroid velocity. Control tests use M9 buffer with no drug. Five to ten worms are tested for each drug concentration and control run. A percent response parameter is calculated by normalizing the centroid velocity at every levamisole concentration to the control runs. These percent response curves are plotted for paper- and plastic-based devices in Fig. 5a and Fig. 5b, respectively. The EC50 values (concentration that produces a response half way between the maximum and minimum response) is 38.46 μM for paper-based devices and is 42.56 μM for plastic-based devices.

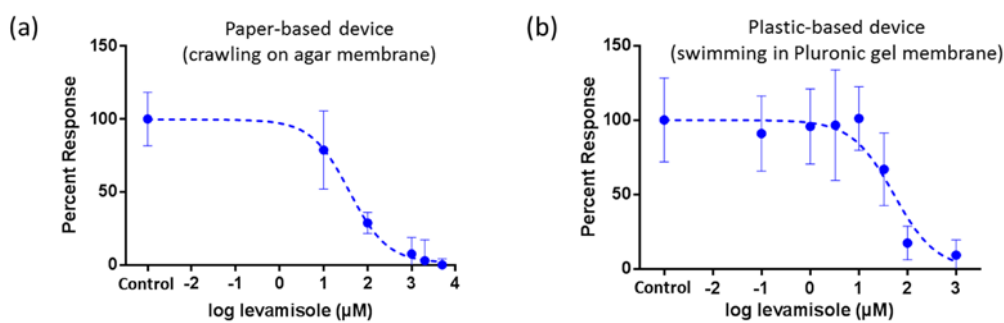

**FIG. 5.** Levamisole drug testing of *C. elegans* on suspended gel membranes. Here a desired concentration of levamisole is prepared in M9 buffer and allowed to permeate into the gel membrane. The centroid velocity of individual worms is tracked at real-time and averaged over the entire recording period. The percentage response is calculated by normalizing the averaged centroid velocities from the experimental tests to those from the control tests. (a) Dose response of *C. elegans* crawling on agar membranes in a paper-based device. (b) Dose response of *C. elegans* swimming in Pluronic gel membranes in a plastic-based device.

**Transfer of worm populations**

The proposed µPADs offer relative ease of worm transfer to agar plates or between membranes. Conventionally, when working with worms on agar plates, the animals are observed under a benchtop microscope and a sterilized wire pick is used to pick and transfer the worms.[30-32] This manual process of handling *C. elegans* requires considerable hand-eye coordination, dexterity, and training. In microfluidic chips, worms are usually accessed in a similar manner through large ports or reservoirs by pipetting or using a wire pick under a microscope.[32,49] With the paper- and plastic-based µPADs discussed here, it is possible to transfer worm populations with minimal to no use of worm picks or a microscope.

Fig. 6 illustrates our methods of transferring worm populations between two membranes. For every step in the figure, a cartoon schematic is on the top and the actual device image is at the bottom. In Fig. 6a(i), two separate paper-based devices are shown in the actual image; the blue membrane represents the fresh device and the red membrane contains a worm population. The paper is simply folded inwards to bring the red membrane in physical contact with the blue membrane (Fig. 6a(ii)). In this instance, the worms are sandwiched in between the two membranes (Fig. 6a(iii)). The paper is then folded outwards, and the worms remain on the bottom blue membrane (Fig. 6a(iv)). By a similar approach shown in Fig. 6b(i-ii), worms are transferred from four plastic-based Pluronic gel membranes (colored with red food dye) to four fresh Pluronic gel membranes (colored with blue food dye). In this case, the two plastic substrates are held such that the red and blue membranes are vertically aligned (Fig. 6b(iii)). Then the plastic substrates are brought in physical contact such that the red and blue membranes mix. The top plastic substrate is now discarded while the worms stay in the membranes of the bottom plastic substrate (Fig. 6b(iv)).

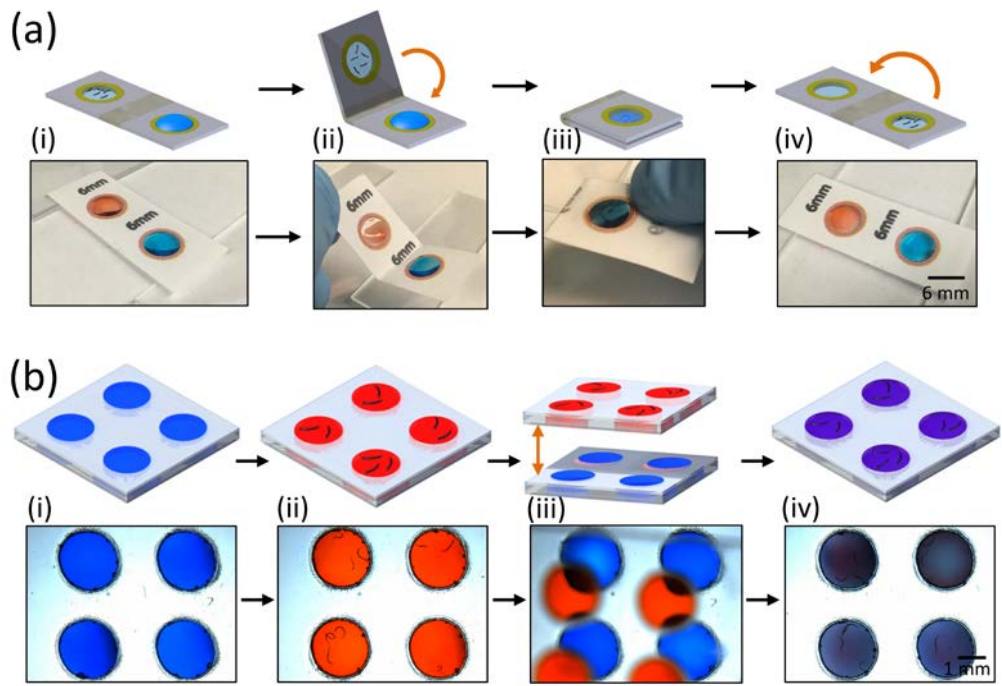

**FIG. 6.** Method to transfer *C. elegans* between two paper or plastic-based µPADs. The worm population is to be transferred from the red-colored membranes to the 'target' blue-colored membranes. (a) To transfer worms between two agar membrane devices, a drop of buffer is added to the target membrane (i). The paper substrate is folded such that red membrane having the worms is brought in contact with the target membrane (ii-iii). After separating the two membranes, worms will have transferred to the target membrane (iv). (b) To transfer worms between two plastic-based devices with Pluronic gel membranes, the top plastic-based device with red membranes (containing the worms) is aligned and lowered onto the bottom device with blue membranes (i-iii) until physical contact is made. The top plastic device is then removed and the worms along with some of the gel have transferred to the bottom plastic device having the blue membrane (iv).

To transfer the worms from a membrane (Fig. 7a(i)) onto an agar plate, the paper-based µPAD is gently lowered on the agar plate as shown in Fig. 7a(ii). Upon lifting the paper substrate, the membrane remains on the agar plate and the worms are able to crawl away freely (Fig. 7a(iii)). Using a similar approach, worms in the suspended Pluronic gel can be transferred to an agar plate. Here the four plastic-based devices having worms (Fig. 7b(i)) are lowered on the agar plate. The Pluronic gel membrane stays on the agar plate while the plastic substrate is discarded (Fig. 7b(ii)). Thereafter worms are free to crawl on the agar plate (Fig. 7b(iii)). In both cases, we are able to transfer all the worms from the membrane devices to the agar plates in five to six independent trials for each device.

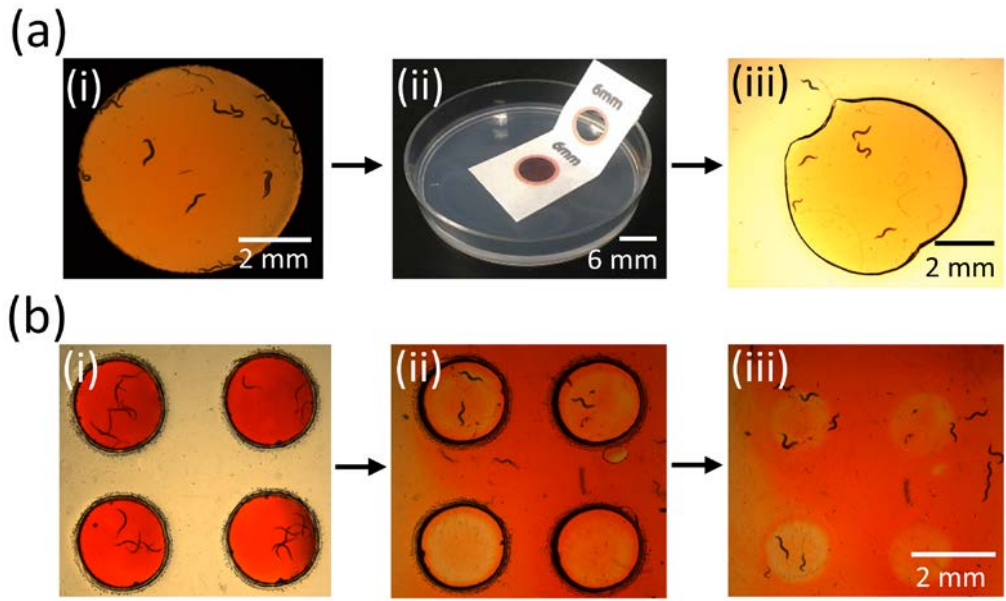

**FIG. 7.** Method to transfer *C. elegans* from paper or plastic-based µPADs to an agar plate. The membranes are colored with a red food dye for visual ease. (a) The paper-based device having a worm population (i) is gently lowered onto the surface of agar plate (ii). When the paper substrate is lifted upwards, its suspended agar membrane detaches and is left on the agar plate (iii). The worms are now free migrate on the agar plate. (b) The Pluronic gel membranes containing worms (i) are lowered onto the agar plate. Once contact is made, the suspended membranes rupture and the worms are transferred to the agar plate (ii). The plastic substrate is removed and the worms are allowed to crawl over the agar plate (iii)

**Imaging multiple worm colonies by z-stacking**

Multilayered paper scaffolds have been used for the *in vitro* co-culture of different cell lines to study the dynamic characteristics of tumor cell migration.[42,43] Using a similar approach, it is possible to image worms on multiple µPADs (each µPAD having worms at a separate experimental condition) that are vertically stacked and aligned together. In Fig. 8a, we stack two Pluronic gel membrane µPADs, one on top of the other (vertical spacing = 2 cm). By manually adjusting the focus of the microscope, we are able to change the focal plane and observe the worm population on the desired plane. Three sample virtual planes are illustrated: yellow, red, and blue planes. In Fig. 8b, two representative worms are highlighted; one lies on the upper device and one on the lower device. When the focus is on the top yellow plane, worm 2 (on the upper device) is clearly visible while worm 1 (on the bottom device) appears blurred. When the red plane is in focus, both worms appear blurred. As the focus is changed to the bottom blue plane, worm 1 is clearly visible while worm 2 appears blurred. The ability to vertically stack multiple planar devices without increasing the spatial footage and image the worm behavior on each membrane independently is not possible on agar plate assays.[32] Such vertical stacking of suspended membranes in microfluidic devices would necessitate multi-step fabrication and bonding techniques, and may not be practical beyond two layers of membranes.[38,41,45]

The cost of consumables in the µPADs is much lower compared to those in agarose plates or polymeric microfluidic devices. For example, using wax printing, it is possible to fabricate approximately 100 paper-based µPADs (4 cm$^2$ of device area) in a 8.5 inch × 11 inch sheet of Whatman Chromatography paper ($57.60 per 100 sheet) using solid ink ($0.0001 per cm$^2$ and assuming 20% ink coverage).[7] This results in ~$0.6 per 100 paper-based µPADs. In a similar manner, approximately 100 plastic-based µPADs (4 cm$^2$ of device area) can be made in a 8.5 inch × 11 inch transparency sheet ($19 per 100 sheet) using adhesive tape ($1.85 per roll having 9.76 m$^2$). The cost is ~$0.15 per 100 plastic-based µPADs. In comparison, the price for Nematode Growth Media (NGM) agarose plates (100 mm) is ~$40.83 per 20 plates from a commercial vendor (Teknova$^{TM}$). The price of consumables for microfluidic chips will be higher and may vary depending on the chip complexity, equipment costs, and labor.

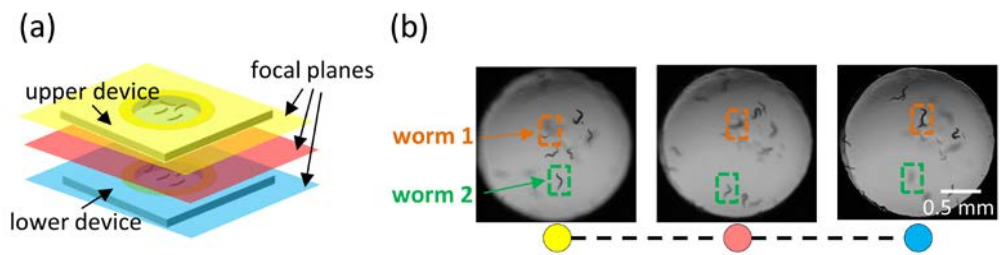

**FIG. 8.** Simultaneous imaging of worms on different focal planes by z-stacking separate µPADs. (a) The membrane devices are vertically stacked (spacing = 2 cm) and aligned to ensure visibility of all membranes. Then by manually adjusting the focus of the microscope, different focal planes can be independently imaged. (b) Images extracted from a video sequence taken as the focal plane of the microscope is changed from the upper membrane (yellow plane) through the middle plane (red plane) to the lower membrane (blue plane). Two representative worms are highlighted where worm 1 is on the lower device and worm 2 is on the upper device. Initially, as we focus on the upper device, worm 2 is clearly visible while worm 1 appears blurred. When we focus to the middle of the two devices, both worms appear blurred. As the focus is now changed to the lower device, 'worm 1' (on the lower device) is clearly visible while 'worm 2' appears blurred.

**CONCLUSION**

In conclusion, paper-based and plastic-based µPADs were developed to facilitate studies on *C. elegans*. The fabrication method required easily-accessible parts (i.e. paper, wax printer, and hot plate) and can thus be leveraged by non-microfluidic laboratories. Compared to agarose plate and polymeric microfluidic assays for worm assays, we showed that the µPADs have advantages such as much lower costs per device, one-step transfer of worm populations from membranes, ability to image multiple z-stacked membranes simultaneously, and easy chemical testing/accessibility due to its open device structure. A Matlab custom tracking program was written to extract the movement parameters of the worms on the agar and Pluronic gel membranes. The tracking program incorporates active contour algorithms to successfully identify suspended membranes from background images, and uses adaptive thresholding algorithms to track worm movement over a time period. We believe the presented μPADs, along with the tracking software, will be appealing for nematode laboratories as a simple, low-cost worm handling, imaging and screening assay.

**ACKNOWLEDGEMENTS**

This work is partially supported by the U.S. National Science Foundation (NSF IDBR-1556370 and CBET-1150867 to S. P.) and Defense Threat Reduction Agency (HDTRA1-15-1-0053 to S. P.). The *C. elegans* culture plates were generously provided by Dr. Jo Anne Powell-Coffman from the Department of Genetics, Development and Cell Biology at Iowa State University.

**REFERENCES**

1. J. Oh and K. Chow, Anal. Methods **7**, 7951 (2015).
2. K. Lee, T. Lee, S. Jeong, H. Choi, N. Heo, J. Park, T. Park and S. Lee, Sensors **12**, 10810 (2012).
3. A. Nilghaz, D. Ballerini and W. Shen, Biomicrofluidics **7**, 051501 (2013).
4. Y. Lu, W. Shi, L. Jiang, J. Qin and B. Lin, Electrophoresis **30**, 1497 (2009).
5. A. Martinez, S. Phillips, M. Butte and G. Whitesides, Angewandte Chemie International Edition **46**, 1318 (2007).
6. D. Cate, J. Adkins, J. Mettakoonpitak and C. Henry, Anal. Chem. **87 (1)**, 19 (2015).
7. E. Carrilho, A. Martinez and G. Whitesides, Anal. Chem. **81**, 7091 (2009).
8. R. Consden, A. Gordon and A. Martin, Biochemical Journal, **38**, 224, (1944).
9. Y. Zhang, C. Zhou, J. Nie, S. Le, Q. Qin, F. Liu, Y. Li and J. Li, Anal. Chem. **86**, 2005 (2014).
10. K. Abe, K. Suzuki and D. Citterio, Anal. Chem. **80**, 6928 (2008).
11. K. Yamada, S. Takaki, N. Komuro, K. Suzuki and D. Citterio, Analyst **139**, 1637 (2014).
12. W. Dungchai, O. Chailapakul and C. Henry, Analyst **136**, 77 (2011).
13. E. Fenton, M. Mascarenas, G. Lopez and S. Sibbett, Applied Materials and Interfaces **1**, 124 (2009).
14. J. Nie, Y. Liang, Y. Zhang, S. Le, D. Li and S. Zhang, Analyst **138**, 671 (2013).
15. J. Olkkonen, K. Lehtinen and T. Erho, Anal. Chem. **82**, 10246 (2010).
16. A. Määttänen, D. Fors, S. Wang, D. Valtakari, P. Ihalainen and J. Peltonen, Sensors and Actuators B:

Chemical **160**, 1404 (2011).

17. D. Liana, B. Raguse, L. Wieczorek, G. Baxter, K. Chuah, J. Gooding and E. Chow, RSC Advances **3**, 8683 (2013).

18. A. Martinez, S. Phillips and G. Whitesides, Proceedings of the National Academy of Sciences of the United States of America **105**, 19606 (2008).

19. H. Liu and R. Crooks, ACS **133**, 17564 (2011).

20. A. Martinez, S. Phillips, Z. Nie, C. Cheng, E. Carrilho, B. Wiley and G. Whitesides, Lab on a Chip **10**, 2499 (2010).

21. E. Fu, B. Lutz, P. Kauffman and P. Yager, Lab Chip **10**, 918 (2010).

22. A. Martinez, S. Phillips, E. Carrilho, S. Thomas, H. Sindi and G. Whitesides, Anal. Chem. **80**, 3699 (2008).

23. A. Yetisen, J. Martinez-Hurtado, A. Garcia-Melendrez, F. da Cruz Vasconcellos and C. Lowe, Sensors and Actuators B: Chemical **196**, 156 (2014).

24. V. Oncescu, D. O'Dell and D. Erickson, Lab on a Chip **13**, 3232 (2013).

25. L. Guan, R. Cao, J. Tian, H. McLiesh, G. Garnier and W. Shen, Cellulose **21**, 717 (2013).

26. L. Shen, J. Hagen and I. Papautsky, Lab on a Chip **12**, 4240 (2012).

27. N. Thom, G. Lewis, K. Yeung and S. Phillips, RSC Advances **4**, 1334 (2014).

28. S. Brenner, Genetics **77**, 71 (1974).

29. A. Corsi, B. Wightman and M. Chalfie, Genetics **200(2)**, 387 (2015).

30. J. Saldanha, S. Pandey and J. Powell-Coffman, Life Sciences in Space Research **10**, 38 (2016).

31. J. Carr, R. Lycke, A. Parashar and S. Pandey, Appl. Phys. Lett. **98 (14)**, 143701 (2011).

32. J. Saldanha, A. Parashar, S. Pandey and J. Powell-Coffman, Toxicol. Sci. **135 (1),** 156 (2013).

33. K. Chung, M. Crane and H. Lu, Nature Methods **5**, 637 (2008).

34. S. Pandey, A. Joseph, R. Lycke and A. Parashar, Advances in Bioscience and Biotechnology **2**, 409 (2011).

35. W. Shi, H. Wen, Y. Lu, Y. Shi, B. Lin and J. Qin, Lab on a Chip **10**, 2855 (2010).

36. G. Aubry, M. Zhan and H. Lu, Lab on a Chip **15(6)**, 1424 (2015).

37. C. Fang-Yen, C. Gabel, A. Samuel, C. Bargmann and L. Avery, Methods in Cell Biology **107**, 177 (2012).

38. N. Chronis, M. Zimmer and C. Bargmann, Nature Methods **4**, 727 (2007).

39. N. Swierczek, A. Giles, C. Rankin and R. Kerr, Nature Methods **8**, 592 (2011).
40. M. Cornaglia, L. Mouchiroud, A. Marette, S. Narasimhan, T. Lehnert, V. Jovaisaite, J. Auwerx and M. Gijs, Scientific Reports **5**, 10192 (2015).
41. B. Gupta and P. Rezai, Micromachines **7(7)**, 123 (2016).
42. R. Derda, S. Tang, A. Laromaine, B. Mosadegh, E. Hong, M. Mwangi, A. Mammoto, D. Ingber and G. M. Whitesides, PloS one **6(5)**, e18940 (2011).
43. G. Camci-Unal, D. Newsome, B. Eustace and G. M. Whitesides, Advanced Healthcare Materials **5**, 641–647 (2016).
44. X. Li, D. Ballerini and W. Shen, Biomicrofluidics **6**, 011301, (2012).
45. S. Hulme and G. Whitesides, Angew. Chem. Int. Ed. Engl. **50**, 4774 (2011).
46. A. Deutmeyer, R. Raman, P. Murphy and S. Pandey, Advances in Bioscience and Biotechnology **2**, 207 (2011).
47. J. Yang, Z. Chen, P. Ching, Q. Shi and X. Li, Lab Chip **13**, 3373 (2013).
48. Y. Zhang, H. Lu, C. Bargmann, Nature **438**, 179 (2005).
49. R. Lycke, A. Parashar and S. Pandey, Biomicrofluidics **7(6)**, 064103 (2013).